\documentclass{appolb}
\usepackage{amsmath,amssymb}
\usepackage{graphicx}
\usepackage{hyperref,doi}
\newcommand{\nc}{\newcommand} 
\nc{\req}[1]{Eq.\,\ref{#1}} 
\nc{\rf}[1]{Fig.~\ref{#1}} 
\nc{\Th}{\ensuremath{T_\mathrm{H}\,}}
\nc{\E}{\mathrm{e}}
\nc{\mydoi}[2]{\href{http://dx.doi.org/#2}{#1}}
\begin{document}
\textwidth=135mm
\textheight=200mm
\title{The Mar(e)k of QGP: Strangeness%
\thanks{Based on two invited lectures (personal and technical) presented at CPOD (Critical Point and Onset of Deconfinement) International Conference held in Wroclaw, Poland May 30th --June 4th, 2016; dedicated to Marek Ga\'zdzicki on occasion of his 60th birthday.}
}
\author{Jan Rafelski
\address{Department of Physics, The University of Arizona\\ Tucson, AZ 85721, USA}
}
\maketitle


\begin{abstract} 
Strangeness signature of of quark-gluon plasma (QGP) is central to the exploration of baryon-dense matter: the search for the critical point and onset of deconfinement.  I report on the discovery of QGP by means of strangeness: The key historical figures and their roles in this quest are introduced and the experimental results obtained are discussed. The important role of antihyperons is emphasized. The statistical hadronization model, and  sudden hadronization are described. Results of present day data analysis:  strangeness and entropy content of a large fireball, and the universal hadronization condition describing key features of all explored collision systems are presented.
\end{abstract}


\section{Introduction}

The introduction in 1964 of the new quark paradigm~\cite{Karl:2010zz} \lq happened\rq\ nearly in parallel to the rise of the thermal model of hadron production precipitated by Hagedorn\rq s invention of the statistical bootstrap model, for a review of Hagedorn\rq s work, see Ref.\,\cite{Rafelski:2016hnq}. The paradigm of quarks explained a large number of properties of elementary particles, allowing at the time one prediction, a new particle, the triple-strange $\Omega^{-}\!(sss)$. On the other hand Hagedorn was following fragmentary experimental data on particle production; these data were not in agreement with the rudimentary statistical particle production models proposed by Koppe~\cite{Koppe:1949zz} and Fermi~\cite{Fermi:1950jd}. Hagedorn\rq s work is a classic example of a discovery case where theory preceded experiment, focusing the direction of future experimental work.

Given the model difficulties that beset interpretation of multi-particle spectra in the early 1960\rq s it would have been easy to abandon the early thermal particle production models in favor of the S-Matrix bootstrap or the Regge-poles, theories hardly anyone reading these pages has ever heard about. Hagedorn persevered, making several visionary contributions that established the statistical-thermal physics in the realm of strong interactions. I believe that his achievements are in comparison to the parallel development of the quark model more imaginative: by trial and error he created a new paradigm {\em before} experimental necessity. His achievement holds, and so today the concept of \lq Hagedorn temperature\rq\ is a part of our vocabulary in the context of particle production. 

About 15 years after the creation of these two great ideas of strong interactions, quarks and Hagedorn\rq s temperature, they merged, giving birth to a new discipline, the physics of quark-gluon plasma (QGP), a new phase of matter; I recently presented my recollection of the related events, this report provides in Sect.~\ref{Dedic} additional information not found in Refs.\,\cite{Rafelski:2016hnq,Rafelski:2015cxa}. The salient feature that matters here is that deconfinement of quarks was not found in the highest energy lepton-hadron collision experiments (recall ideas about \lq asymptotic freedom\rq), but in domains of space-time where the aether of modern day, the structured quantum-vacuum, is dissolved by the collective action of many participating nucleons. 

This simple insight is again paradigm-setting, opening the door to the study of the color deconfined QGP in laboratory relativistic heavy-ion collision experiments. And yet scant attention is paid to this fascinating physics insight: in current US funding agency language we hear about the \lq energy\rq\ and the \lq intensity\rq\ frontiers; meaning that to make further discoveries we need to achieve highest elementary particle beam energy or highest particle beam intensity. By saying this we miss out on the Hagedorn temperature frontier; that is, the exploration in the laboratory of the conditions last seen in the early Universe at an age of about 20\;$\mu$s.

In the following I first look back at Marek\rq s contribution to the physics of strangeness signature of QGP and the QGP discovery. In my opinion, Marek is among the few relevant personalities in this field. I will make my case why, in my mind, there is little doubt that Marek initiated the experimental effort to discover QGP through its strangeness signature~\cite{Anikina:1984zh}. In his work Marek was in the first wave of interest that followed the appearance of my fledgling proposals about the strangeness signature of QGP, see for example Ref.\,\cite{Rafelski:1980fy}. Another important component of this report showing Marek\rq s achievements is the brief history of Marek and my personal interaction, presented against the background of the rapidly evolving experimental and science-political situation. In this context I update below Ref.\,\cite{Rafelski:2015cxa} describing and explaining to the end the causes of the divergent interpretation of research results that seeded confusion in the field for many years. I also describe events that in my opinion have led to the creation of the unicorn view of QGP physics. 

The pivotal CERN experiments, which to the disbelief of some of the discoverers (see Sect.~\ref{ref:dst}) created the QGP phase of matter, were performed by the end of the last century. The QGP discovery was reconfirmed within 5 years by work done at the relativistic heavy-ion collider (RHIC) at the BNL laboratory in US. However, books addressing particle and/or nuclear physics written since the QGP discovery prefer not to speak of QGP. If described at all I see QGP addressed just like one writes about unicorns. With QGP discovery topic remaining in eyes of many unsettled, just about anyone working on the subject today thinks s/he contributed decisively; QGP is truly a hot topic -- and that is good for the field as there are many young faces we see at conferences! However, this continued lack of clarity about prediction and experimental results is in Marek\rq s and my view the origin of the ongoing lack of consensus about the experimental QGP discovery.

I present in a historical Sect.\,\ref{Dedic} the scientific context of the present day: the early beginning and evolving experimental research program at CERN in Sect.\,\ref{ref:dst}, the first sighting of strangeness from QGP in Sect.\,\ref{ssec:discovery}, the wealth of present day experimental results in Sect.\,\ref{ssec:CERNQGP} that prompted the CERN QGP discovery announcement, and describe the  present day status in Sect.\,\ref{ssec:horn}. The NA61/SHINE experiment is the \lq icing on the cake\rq\ in the CERN-QGP discovery story. Everyone knows that without Marek there would have been no NA61/SHINE; he had the idea of merging several constituencies; he was able to get a large CERN experiment going in a time that all resources were focused on the LHC. Marek managed to develop a research program that complements and in fact outclasses in terms of data precision till this day the effort of an entire large laboratory, BNL, where the RHIC Beam Energy Scan (BES) project is sited. Bravo! 

In the following Sect.\,\ref{sec:SHM} the statistical hadronization model used in the experimental data analysis is introduced. I present in Sect.\,\ref{ssec:fire}  sudden hadronization,  and show some important analysis results in the following Sect.\,\ref{ssec:analysis}: focusing on strangeness and entropy, and the universality of hadronization.  This report closes with a few forward-looking remarks in Sect.\,\ref{sec:conc}.
\section{The Mar(e)k of $s$}\label{Dedic}
\subsection{The beginning}\label{ref:dst}
Over the past 30 years of friendship and collaboration, I have been privileged to work with Marek; our science-lines first intersected around 1982. In an experimental report on production of strange hadrons~\cite{Anikina:1984zh}, we see in Ref.~[2] of this manuscript a mention of the second wave of theoretical strangeness production, work I was doing in 1981/1982, UFTP preprints 80/82~\cite{Rafelski:1982ii} and 86/82~\cite{Rafelski:1982bi}. In the body of this arguably first experimental work exploring strangeness production and seeking its enhancement in relativistic heavy-ion collisions, we read in the pointer to this Ref.~[2]: \lq\lq According to the existing theoretical considerations one can expect in particular an enhancement of strange-particle production\ldots\;.\rq\rq\ 

The work of Marek that followed supports the conjecture that he spearheaded the Dubna group interest and insight into strangeness. Following his Dubna debut he was able to join the CERN-NA35 collaboration where he developed the strangeness signature of QGP. NA35 was a scion of the LBNL-GSI BEVELAC experimental effort with LBNL\rq s Howell Pugh being the main administrative and intellectual force for strangeness, as my personal correspondence from this period shows. Howell was a key member of both the NA35 and the NA36 experiments; however, NA36 had instrumental difficulties. While initially the objective of NA35 was the exploration of equations of state of dense nuclear matter, which was a direct continuation of the effort carried out at LBNL-BEVALAC, Marek\rq s arrival, and the fade-out of NA36 presented the NA35 experimental program with the opportunity to enter forcefully into a novel domain of heavy-ion physics, the strangeness signature of QGP. But as the following events show, not everyone in the collaboration was ready to move in this direction immediately.
 
Writing in year 2000~\cite{Odyniec:2001}, Gra\.zyna Odyniec of LBNL largely corroborates my memory and the contents of my correspondence: \lq\lq From the very beginning Howell (Pugh), with firmness and clarity, advocated the study of strange baryon and antibaryon production. He played a leading role in launching two of the major CERN heavy-ion experiments: NA35 and NA36, the latter being exclusively dedicated to measurements of hyperons. Strangeness enhancement predicted by theorists was discovered by NA35 and reported at the Quark Matter Conference in 1988.\rq\rq\ I have reviewed the proceedings of this meeting held in September 1988 and while Marek~\cite{Gazdzicki:1989kd} presents NA35 strangeness $\Lambda$ and $K^0$ enhancement results in S-S collisions, everything shown in the proceedings of the meeting published July 1989 is overprinted with the warning, \lq preliminary\rq. 

These results are presented in their extended and final format two years later~\cite{Bartke:1990cn} (submitted 17 April 1990; in revised form 2 July 1990). As Marek remarked in a personal conversation, the title of this paper should have been \lq\lq Strangeness Enhancement as the Evidence of QGP,\rq\rq\ but both the collaboration dynamics (I count 67 authors on this manuscript, a large number as measured by the norm of the time period) and the review process prevented such a claim. Thus instead the manuscript has the non-telling title, \lq\lq Neutral strange particle production in sulphur-sulphur and proton-sulphur collisions at 200 GeV/nucleon.\rq\rq\ The abstract states, \lq\lq Significant enhancement of the multiplicities of all observed strange particles relative to negative hadrons was observed in central S-S collisions, as compared to $p+p$ and $p+$S collisions.\rq\rq\ In the concluding section, buried in a lot of ink, one finds, \lq\lq Thus our observation \ldots appears to be consistent with a dynamical evolution that passes through a deconfinement stage.\rq\rq\ 

Had the NA35 collaboration ended the concluding section here one could rightly see them agreeing with Marek and claiming discovery of QGP as well. Reading on in Ref.\,\cite{Bartke:1990cn} we see: \lq\lq However (according to) [11, 14] this may not be the only explanation because the possible pre-equilibrium aspects of the early interpenetration stage, or even the conceivable overall off-equilibrium nature of the entire dynamics, may also lead to enhanced strangeness production, even without plasma formation.\rq\rq\ So what worry did NA35 have? Let us here look at these two references: a) I believe the unpublished Frankfurt preprint [11 from Ref.\,\cite{Bartke:1990cn}] was later recognized as a collection of errors and omissions to be rectified years later by Stefan Bass (apologies if you disagree); b) Preprint and paper [14 from Ref.\,\cite{Bartke:1990cn}] were immaterial as in an earlier work together with a very talented student Peter Koch I had shown that evolution within hadronic gas could not chemically equilibrate strangeness~\cite{Koch:1984tz}. In any case, after making a step in the right direction, in the end NA35 decided in their initial flagship S-S publication not to claim directly or indirectly that they were seeing QGP.

I know that Marek has fought bravely to push strangeness and strange antihyperons within the NA35 collaboration as a QGP story and that he was sure of his raw data analysis results long before they were checked and cross checked by others in his collaboration. However, in the early 1990\rq s, Marek was still a young experimentalist from an odd place in the East, surrounded by Western senior scientists who did not want to believe what they saw as result of their experiment. 

Searching for definitive NA35 words on strangeness enhancement I came across the QM1990 conference report based on a presentation in mid-May 1990 by the spokesman of NA35 Reinhard Stock, printed in April 1991~\cite{Baechler:1991pp}. In this work one reads: \lq\lq In a previous NA35 experiment we reported [4] results for central $^{16}$O+Au collisions which did not exhibit spectacular (strangeness) enhancements over the corresponding $p$+Au data.\rq\rq\ Thus in the opinion of Reinhard Stock, the discovery of strangeness enhancement is in the S-S collisions he reports on. Reinhard concludes in his report: \lq\lq we have demonstrated a two-fold increase in the relative $s + \bar s$ concentration in central S-S collisions, both as reflected in the $K/\pi$ ratio and in the hyperon multiplicities. A final explanation in terms of reaction dynamics has not been given as of yet.\rq\rq\ We see that in Summer 1990 and even later the NA35 collaboration does not want as a group to introduce the QGP interpretation of the strangeness enhancement results, even though \lq \ldots  a two-fold increase in the relative $s + \bar s$ concentration \ldots\rq\ was expected based on QGP dynamics~\cite{Koch:1986ud}.

That is in retrospect a case of bad judgment by the NA35 collaboration. Another experiment takes under the leadership of Emanuele Quercigh the center stage of strangeness production and QGP search with results on: $\Lambda$ and $\bar{\Lambda}$~\cite{Abatzis:1990cm} (a CERN preprint of 18 April 1990); on $\Xi^-$, $\overline{\Xi^-}$~\cite{Abatzis:1990gz} (a CERN preprint of 8 November 1990); and a systematic exploration of QGP characteristic behavior for both~\cite{Abatzis:1991ju} (a CERN preprint of 5 July 1991). Here the WA85 collaboration takes a firm position in favor of QGP discovery with the words: \lq\lq The(se) results indicate that our $\overline{\Xi^-}$ production rate, relative to $\bar{\Lambda}$, is enhanced with respect to $pp$ interactions; this result is difficult to explain in terms of non-QGP models [11] or QGP models with complete hadronization dynamics [12]. We note, however, that sudden hadronization from QGP near equilibrium could reproduce this enhancement~[2].\rq\rq\ Ref.\,[2 from Ref.\,~\cite{Abatzis:1991ju}] is my work~\cite{Rafelski:1991rh} published in March 1991; more about this below. 

I interpret the sequence of events that transpired now more than 27 years ago as follows: Marek was not allowed by NA35 to go on record as having discovered a) strangeness enhancement in heavy-ion collisions; or b) to have discovered quark-gluon plasma itself. NA35 ducks the QGP question again in 1990 and Reinhard speaking and writing for NA35 did not in 1990/91 go on record with anything QGP-like. In fact the NA35 collaboration continued to disbelieve its own results for another year or longer. I talked to Marek about this and I can say he is more than less in agreement with my views.

I see in NA35 Ref.\,\cite{Bachler:1992js} (at CERN preprint server in October 1992) the first inkling that the internal collaboration dynamics is evolving to accept strangeness enhancement, but not yet QGP discovery. In conclusion section of Ref.\,\cite{Bachler:1992js} we read \lq\lq 4. Neither the FRITIOF nor the VENUS model gives a satisfactory description of the full set of the results\ldots\rq\rq\ and, just below, backtracking a bit, \lq\lq 5. S-S data extrapolated to the full phase space show that the observed strangeness enhancement appears mainly as kaon-hyperon pairs which indicates that this enhancement comes from the region of nonzero baryochemical potential.\rq\rq\ I find no mention of QGP in this NA35 comprehensive 1992 report other than in the first phrase of the manuscript, a comment introduced as a motivation for this experimental work. 

Thus we see that even though Marek, according to my memory was personally confident of his results and understood how these allowed to claim QGP for NA35, at least for a year after WA85 took the position that its data were QGP related, NA35 remained on the sidelines of the QGP discovery story. Finally, when the tide in the collaboration began to turn towards accepting the possibility if not the fact that NA35 had made an extraordinary QGP discovery, Reinhard wrote an open letter to his collaboration complaining that NA35 results were going unnoticed. Instead of blaming his own colleagues he blamed for this situation a lamp post, which happened to be me, yes, me. I still have his letter (which I think I was not supposed to see but which ended on my desk coming from several directions) and our correspondence that ensued. The point in the matter is that NA35 damaged its credibility by dancing around rather than making a claim that was of consequence. The CERN $\Omega\rq$ spectrometer hyperon experiment WA85 reported already in 1990/91 and in a definitive manner a series of perplexing strangeness hyperon and in particular antihyperon results, giving these results a QGP discovery tilt. 

\subsection{Finding quark-gluon plasma at CERN}\label{ssec:discovery}
A first theoretical analysis of the strangeness CERN $\Omega\rq$ spectrometer experiment WA85 became possible in late 1990 and was presented in February 1991 at a week-long workshop at CERN organized by Helmut Satz; it was published soon after~\cite{Rafelski:1991rh}. In this work strange baryon and antibaryon particle production data for S-W collisions were used to determine the \lq chemical\rq\ properties of the particle source, {\it i.e.} the chemical potentials and phase space occupancy, topics that we return to in Sect.~\ref{sec:SHM}. This paper marked the beginning of the development of the statistical hadronization model (SHM), the present day \lq gold\rq\ standard in the study of hadronization of QGP. In collaboration with my friend Jean Letessier the model was completed, allowing many analysis results to be published in 1993/94.

I open the abstract of the 1991 analysis with the words: \lq\lq Experimental results on strange anti-baryon production in nuclear S-W collisions at $200 A$\,GeV are described in terms of a simple model of an explosively disintegrating quark-gluon plasma (QGP).\rq\rq\ In conclusion I close with, \lq\lq We have presented here a method and provided a wealth of detailed predictions, which may be employed to study the evidence for the QGP origin of high $p_\bot$ strange baryons and anti-baryons.\rq\rq\ The paper I cited above~\cite{Abatzis:1991ju} echoes this point of view, taking the WA85 collaboration in 1991 to the cliff. Today, we can say that with this 1990/91 analysis method and the WA85 results and claims of the period, the QGP was discovered. More on this is also seen in a popular review I presented with the spokesman of WA85 Emanuele Quercigh shortly after the CERN announced (February 2000) QGP discovery, see Ref.\,\cite{QR2000}.

Seeing this strangeness-WA85 CERN experiment analysis~\cite{Rafelski:1991rh}, Marek lobbied me with an inviting remark that today still reverberates in my memory, \lq\lq \ldots would it not be nice to also apply these methods to other experiments?\rq\rq\ We began the discussion of the data available in NA35. The situation was not all good: we needed lots of data for the rudimentary SHM to be useful; NA35 was using a photographic method based on triggered events in a streamer chamber, requiring human photograph scanning of visible tracks. This was a very time intensive process and when track density was high, this approach was inefficient. All this meant that our initial objective, the confirmation study of the equivalent to WA85 S-Pb reactions, was not possible. However, we soon realized that the S-S experimental results were both sufficiently precise and rich in particles considered, and therefore could be analyzed. Our discussions resulted in an analysis publication of the NA35 S-S $200 A$ GeV collision results~\cite{Sollfrank:1993wn} (submitted in August 1993).

This project was carried out jointly with a very young student, Josef Sollfrank. His thesis advisor Ulrich Heinz made an \lq improvement\rq\ in the draft manuscript by removing every mention of quark-gluon plasma from the entire paper. We thus read in the conclusions, \lq\lq This agrees with the notion of common chemical and thermal freeze-out following explosive disintegration of a high entropy source,\ldots\rq\rq; here QGP=high entropy source. Just like the early WA85 analysis, the NA35 S-S $200 A$ GeV analysis was consistent with our predictions about strangeness production in QGP. However, NA35 did not yet have multi-strange particles which I always viewed to be the unique QGP signature, immune to reinterpretations. At the time multi-strange hyperons were only in the hands of WA85.

\begin{figure}[bht]
\centerline{%
\includegraphics[width=12.cm]{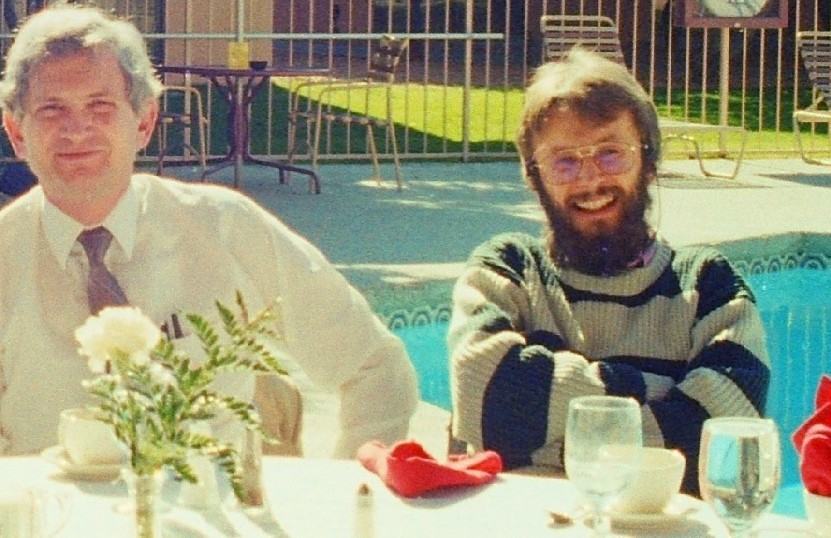}}
\caption{Marek Ga\'zdzicki (off-center right) in mid-January 1995 relaxing with the author pool-side at the SQM-1995 conference in Tucson, Arizona~\cite{Rafelski:1995zq}.}
\label{MG_SQM}
\end{figure}

In early 1995 strangeness enthusiasts celebrated the discovery of the QGP, a new phase of matter at a meeting in Tucson~\cite{Rafelski:1995zq}; a poolside working group is seen in \rf{MG_SQM}. In the following five years the continued measurements of the yields of multi-strange (anti)baryons by the $\Omega\rq$-spectrometer experiments WA94 \& WA97, both evolving in sequence from WA85, and the parallel work of Marek within NA49 (successor to NA35) in my opinion sealed the QGP discovery case. The observed production yields of antihyperons, the signature of QGP put forward in my early work~\cite{Rafelski:1980fy,Rafelski:1982ii}, could not be matched in now mature microscopic hadron collision simulations. 

However, this situation did not last long: people are inventive and soon new models arose that could produce whatever was needed to invalidate the strange signature of QGP. Here it is important to remember that newly invent effects must apply to all data, and that inventions of exotic mechanisms after data is known is not as difficult or uncertain as a prediction of the experimental outcome.

On the topic of anti-hyperons in NA35: Reinhard Stock shows $\overline{\Lambda}$ for $p_\bot>0.5$\,GeV at QM1990  meeting, and we find this picture in the NA35 publication~\cite{Bachler:1992js}. A full $4\pi$ result appears in Summer 1994~\cite{Alber:1994tz}, and in July 1995 a direct comparison with $pp$ reactions is presented~\cite{Foka:1995Thesis} (see Fig.~8.24, p.271 shown below, \rf{AntiHypFig}). In discussion of this figure we read (p.268): \lq\lq The enhancement (of $\overline{\Lambda}$) at mid-rapidity is a factor 6 in $^{32}$S-S\ldots strange particle production that is not (due to) a simple superposition of elementary interactions.\rq\rq\ Translating from the French thesis resume: \lq\lq The question if we can conclude that QGP has been observed is the topic of hot debates and this should be considered within the context of many other observables.\rq\rq\ It would have been so much more interesting if these words had instead been, \lq These results agree with antihyperon enhancement observed by WA85/94. The antihyperon signature of QGP has been observed and confirmed.\rq

\begin{figure}[bht]
\centerline{%
\includegraphics[width=6cm]{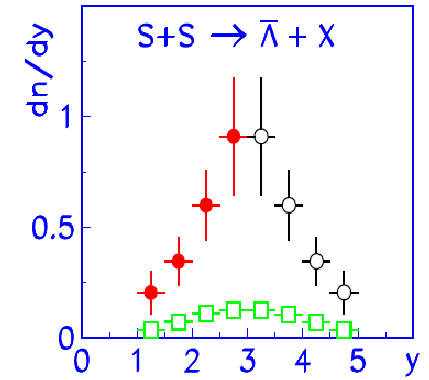} 
\includegraphics[width=5.8cm]{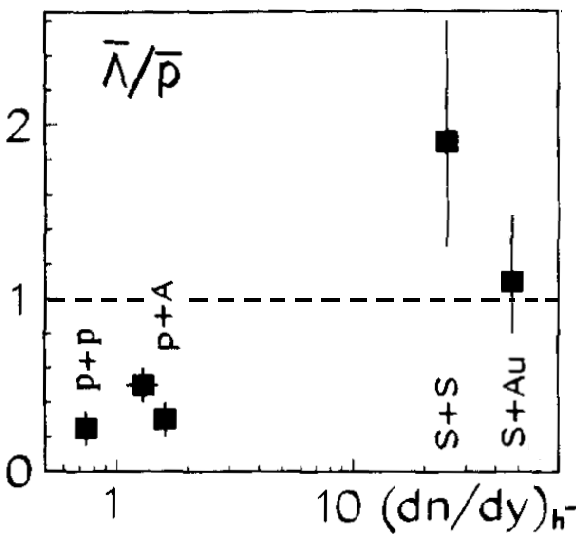}}
\caption{NA35-antihyperon results for 200 GeV/A S-A collisions of July 1995. Left: Distribution in rapidity of $\bar\Lambda$ for S-S compared to the yield in $pp$ collisions scaled up with relative abundance of negatives $h^-$, from Y. Foka thesis~\cite{Foka:1995Thesis}; Right: $\bar\Lambda/\bar p$ ratio as a function of mean $h^-$ multiplicity. A ratio near and above unity was the 1980 predicted QGP signature~\cite{Rafelski:1980fy}, adapted from Ref.\,\cite{Alber:1996mq}.}
\label{AntiHypFig}
\end{figure}

The NA35 presented the ratio $ \overline{\Lambda}/ \bar p\lesssim 1.4$ measured near mid-rapidity in Summer 1995~\cite{Alber:1996mq}, showing enhancement by a factor 3 to 5, dependent on the collision system as compared to measurement in more elementary reactions. This was the QGP signature of my first strangeness papers in 1980~\cite{Rafelski:1980fy}. Due to the shift of the \lq central\rq\ rapidity for asymmetric collisions the decrease in this ratio as the asymmetry increases is in agreement with theoretical expectation, these results are seen on right in \rf{AntiHypFig}. In presenting this result, NA35 pulled equal with the work of the  WA85/94 strangeness group focused on multi-strange ratios, for $ \overline{\Xi}/\overline{\Lambda}$ see \eg\ the 1993 review of David Evans~\cite{Evans:1994sg}. A full summary of all results is seen in the review of Federico Antinori of 1997~\cite{Antinori:1997nn} shown in \rf{WA85WA94AntiHFig} with data referring to the WA85/94 reports presented in January 1995 at QM1995~\cite{DiBari:1995cy,Kinson:1995cz}. 

\begin{figure}[bht]
\centerline{%
\includegraphics[width=7.4cm]{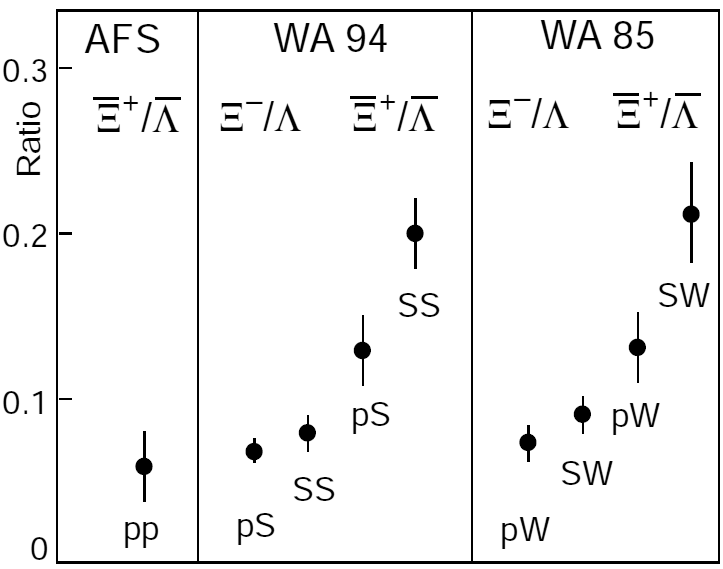}}
\caption{WA85 and WA94 hyperon and antihyperon results for 200 GeV/A S-A collisions of January 1995, adopted from~\cite{Antinori:1997nn}. }
\label{WA85WA94AntiHFig}
\end{figure}

\subsection{The CERN QGP discovery announcement follow-up}\label{ssec:CERNQGP}
For many years that followed Marek and I worked to advance the common cause of the strangeness signature of QGP. An example is access to experimental data. Initially just for me, but soon for his many other colleagues and friends, Marek created an informal compendium of all relevant experimental results emanating from the NA35/NA49 experiment. This document evolved, being updated and circulated, as Marek and his students made a great effort to create a reliable, consistent, and early available data reference. 

As the analysis of the growing body of data improved it was hard for Marek and me to have any doubt about the fact that QGP had been discovered by strangeness and strange antibaryon signature alone. As noted already, the $\Omega\rq$-spectrometer experimental series WA85/WA94/WA97 evolving into NA57 had a similar physics view. The two experimental families made a common effort to achieve mutual understanding of all results. In the totality of results that were presented I see demonstration of: i) thermal production of strangeness flavor; ii) free motion of quarks in deconfined region; and iii) recombinant production of strange antihyperons. This sequence of events is at the origin of the antihyperon QGP signature.

Seeing the CERN S-W/Pb results seen in Sect.\,\ref{ssec:discovery} we could not but be convinced that than forthcoming Pb-Pb experiments at CERN should confirm and cement the strangeness based QGP discovery in the old millennium. On this note let me state that in my opinion the CERN announcement of QGP made in February 2000 was unnecessarily diluted by the inclusion of consensus-building wide and diverse heavy-ion physics results. By making the decision to seek \lq approval\rq\ and consensus, CERN included experiments that offered non-convincing, non-relevant, and even maybe outright wrong results. Agreeing to this procedure CERN opened the QGP discovery to much criticism where both interpretation of data and result validity were questioned. In my view had CERN stuck to its flagship strangeness signature, at first some consternation may have emanated from those working on other QGP signatures, but there would have been few if any doubts possible about the actual experimental results, while with time theoretical objections could and would have been cleaned out.

The situation at CERN was created as I believe by one of the two co-authors of the so-called CERN \lq consensus\rq\ report~\cite{Heinz:2000bk} signed by Ulrich Heinz and Maurice Jacob. This is the same Ulrich Heinz who deleted the mention of QGP in 1993 in our joint publication~\cite{Sollfrank:1993wn}. I later heard from Maurice Jacob that Ulrich Heinz wrote this report. Maurice made some improvements but did not materially impact the consensus Ulrich Heinz created working with all experimental groups. This is corroborated by my personal memory and other written records. For example, shortly before the CERN announcement of QGP I was making an effort to publish a paper, Ref.\,\cite{Rafelski:1999xv} in {\it The Physical Review Letters} (PRL) where at the time Ulrich Heinz was Divisional Associate Editor (1998-2000). Therefore I discussed with him the manuscript contents several times and still have our extensive correspondence.

In the manuscript~\cite{Rafelski:1999xv} Jean Letessier and I showed that the freeze-out of hadron abundance was \lq out of chemical equilibrium\rq\ at $T\lesssim 145$\,MeV. On the other hand, Ulrich Heinz was instead writing, verbatim from Ref.\,\cite{Heinz:2000bk} \lq\lq The theoretical analysis of the measured hadron abundances (NA44, NA45, NA49, NA50, NA52, WA97, WA98) shows that they reflect a state of \lq chemical equilibrium\rq\ at a temperature of about 170 MeV.\rq\rq\ I explained how this false hadronization point was created in Ref.\,\cite{Rafelski:2015cxa}, see Sect.~10.3\;: \lq 170 MeV\rq\ was prompted by off-the-record lattice comments that were seconded by analysis carried out in a simplified SHM model context.

I am convinced today that Ulrich Heinz recognized that the chemical non-equilibrium description of hadron freeze-out~\cite{Rafelski:1999xv} was correct and that strangeness had locked in the QGP discovery. However, he was able to create enough confusion to blur the public understanding. By multitude of his actions beginning in 1993 and as I believe continuing today he has precluded recognition of one of the most brilliant pieces of experimental work, the discovery by strangeness signature of QGP at CERN. In this circumstance the RHIC community made its own QGP discovery case~\cite{RHIC2005a,RHIC2005b,RHIC2005c,RHIC2005d} five years after CERN.

\begin{figure}[bht]
\centerline{%
\includegraphics[width=8.8cm]{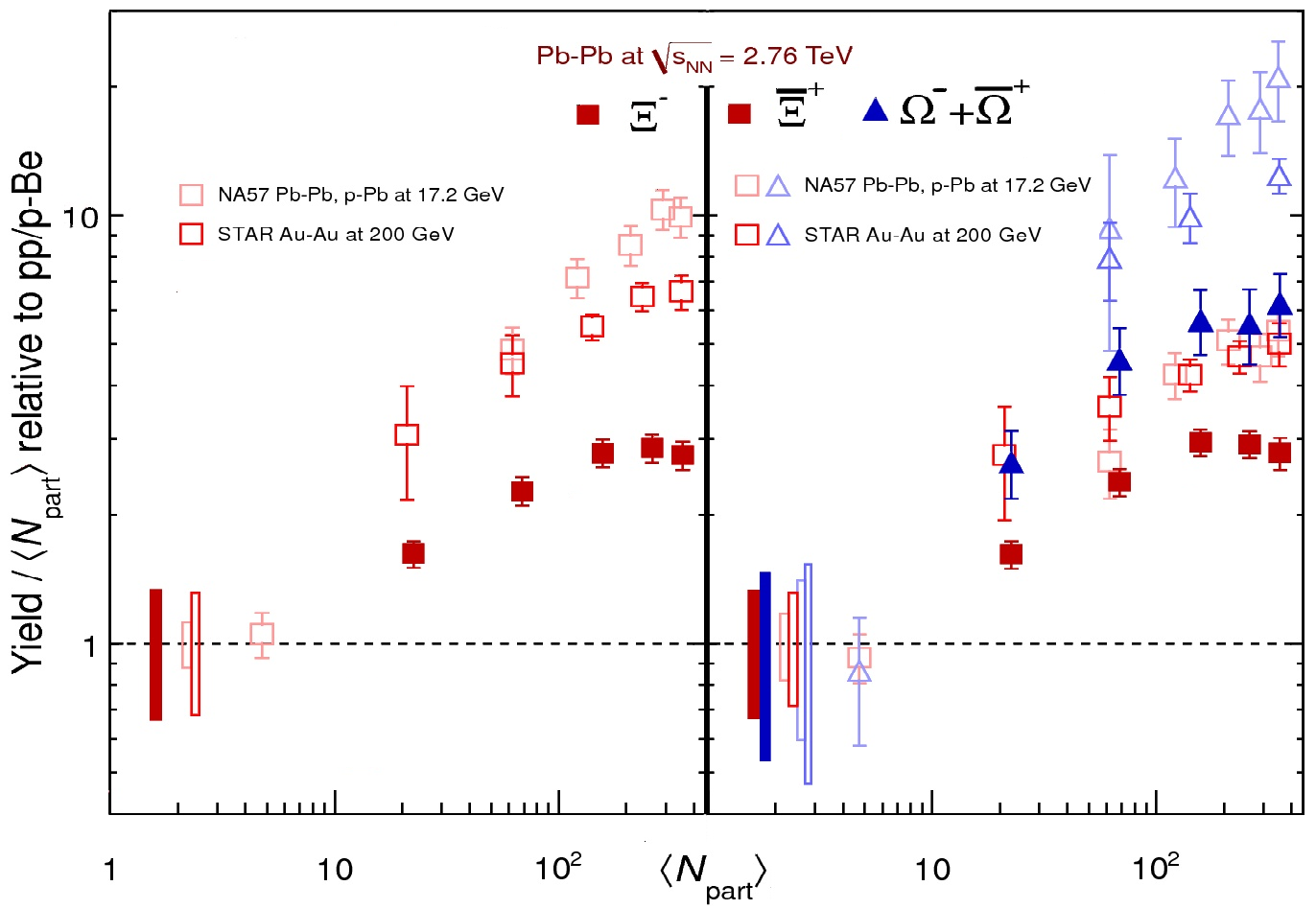}}
\caption{ALICE, STAR, and NA57 $N_\mathrm{part}$ normalized hyperon (left), and antihyperon (right) yields as function of $N_\mathrm{part}$; yields renormalized arbitrarily to unity for smallest $N_\mathrm{part}$ available, adopted from~\cite{ABELEV:2013zaa}.}
\label{AliceALL}
\end{figure}

The experimental situation is today in my view overwhelmingly in favor of QGP interpretation of the hyperon signature. In \rf{AliceALL} we see the current status of hyperon and antihyperon production. The yields divided by the number of participants $N_\mathrm{part}$ are shown as function of collision centrality expressed in terms of $N_\mathrm{part}$ for three different energy domains, ALICE, RHIC and SPS. On the left we see hyperons and on the right antihyperons; in both cases the yield is on logarithmic scale. We see enhancement of up to factor 20 -- the hierarchy and magnitude of enhancements is as I predicted in 1980\rq s, at first alone and later elaborating all details in collaboration with Peter Koch and Berndt M\"uller~\cite{Koch:1986ud}; moreover it is worth noting that the hyperon yield scaling is by far the strongest medium effect observed in all of relativistic heavy-ion collision experiments. Note further that since the absolute yield is in \rf{AliceALL} arbitrarily normalized to the smallest experimental result available in terms of $N_\mathrm{part}$, these curves do not overlay as well as they could.

\begin{figure}[bht]
\centerline{%
\includegraphics[width=6.5cm]{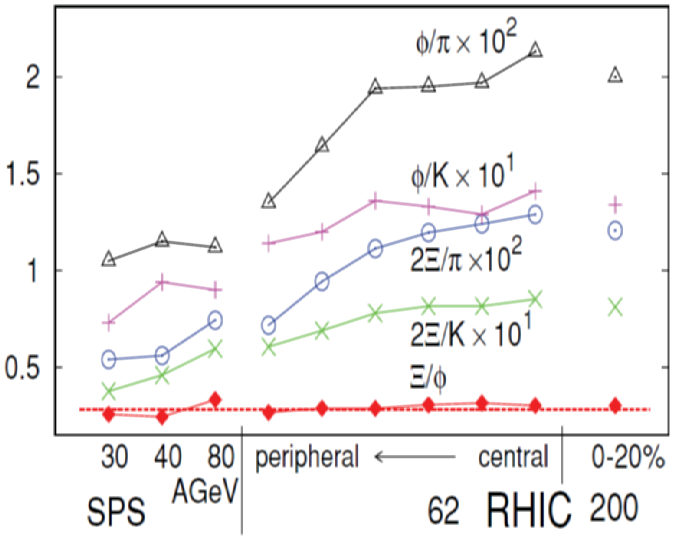}
\includegraphics[width=5.6cm]{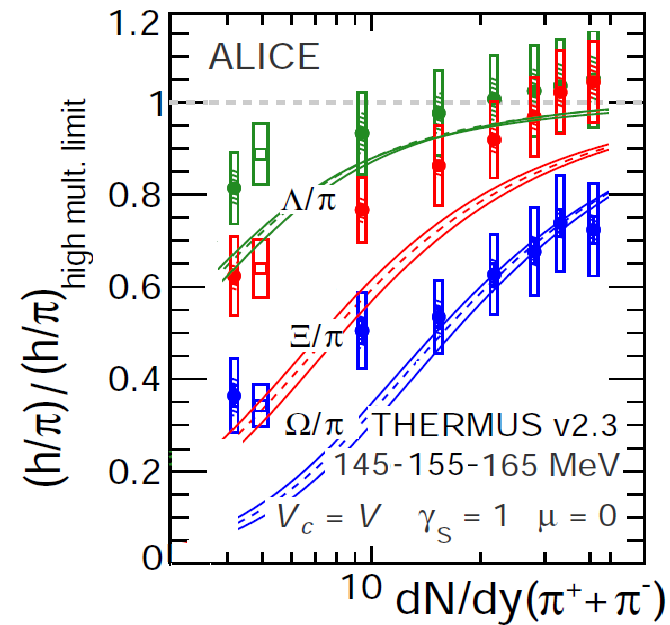}
}
\caption{\underline{Left:}Data points of $\Xi/\phi$ (bottom red diamonds) and a straight line at 0.281; compared to ratios of these particles with $\pi$ and K. Adapted from Ref.\;\cite{Petran:2009dc}. \underline{Right:} ALICE $p$A (filled circles), $pp$ (open squares) hyperon to pion ($\Lambda/\pi$ in green, $\Xi/\pi$ in red, and $\Omega/\pi$ in blue) ratios as a function of charged pion yields. The $h/\pi$ data are the ratios of the particle and antiparticle sums, except for the $2\Lambda/(\pi^-+\pi^+)$ data points. Data compared to a canonical phase space model, with values (arbitrarily) normalized to the high multiplicity limit, which is given by the mean of the 0-60\% highest multiplicity Pb-Pb measurements for the data and by the grand canonical limit for the model bands. Adapted from Ref.\;~\cite{Adam:2015vsf}.}
\label{CanonicalEndFig}
\end{figure}

Are there alternatives to QGP interpretation? My friend Krzysztof Redlich argued that the enhancement could be a mirage created by the varying strangeness suppression in the small collision system reference data, this is implemented in the \rq canonical strangeness suppression\rq\ model. Naturally, to some degree this effect exists but only when presenting results along the method of \rf{AliceALL} as I described in 1980 in the study of volume dependence of the canonical phase space, for full discussion see my SQM2001 review~\cite{Rafelski:2001bu}. To appreciate the relevance of the canonical model we note that it predicts that the production of $\Xi(ssq)$ must differ profoundly from that of $\phi(s\bar s)$ which is volume independent while in QGP breakup model both yields track each other as $s$ and $\bar s$ yields are equivalent. 

In \rf{CanonicalEndFig}  we see a study of multi-strange particles~\cite{Petran:2009dc}, here in particular the $\phi(s\bar s)$ and $\Xi(ssq),\overline{\Xi}(\bar s\bar s\bar q)$, where the geometric average $\Xi=\sqrt{\Xi(ssq)\overline{\Xi}(\bar s\bar s\bar q)}$ (and similarly for kaons and pions) is employed to avoid chemical potential dependence arising significantly for SPS data. The $\Xi/\phi$ ratio is found to be a constant, see the straight line at $\Xi/\phi=0.281$. This shows that in a wide range of collision centrality and energy the mechanism of production of these multi-strange particles behaves as is expected in a QGP breakup according to grand-canonical phase space, ruling out the canonical model. 

In addition on the right in \rf{CanonicalEndFig} we see ALICE~\cite{Adam:2015vsf}  modeling the  Krzysztof idea that there is no hyperon production enhancement in highest pion multiplicity  $p$A and $pp$ events, but a varying degree of suppression in the low multiplicity data. Unlike in \rf{AliceALL} now all experimental results are (arbitrarily) normalized to the high multiplicity limit. The experimental yields are compared to the canonical phase space with a unit system size $V_0$  and temperatures $T=145,155,165$  shown as bands. Chemical equilibrium is assumed $\gamma_q=\gamma_s=1$.  We see a disagreement by up to 5 s.d. in the low charge multiplicity class of events; note that thee discrepancy is highly significant for $\Omega$ (in blue), and $\Xi$ (in red) since the model is geared by volume parameter choice to fit best the $\Lambda$ (in green) behavior, and even in this one functional dependence the canonical model performs poorly.

I believe that at the time of writing the only explanation of the  (multi-strange) (anti) hyperon production in relativistic heavy-ion collisions is in terms of a QGP sudden disintegration model, proposed in 1990/91~\cite{Rafelski:1991rh} completing the mechanism of enhancement predicted in 1980~\cite{Rafelski:1980fy} and allowing to explain and model the first WA85 antihyperon results~\cite{Abatzis:1991ju}.

\subsection{The Horn and the Hadronization Temperature}\label{ssec:horn}
What happened since the CERN QGP announcement event of 10 February 2000? We have seen beautiful LHC strangeness results presented by the ALICE collaboration, a topic of significance as the Universality of physics of QGP formation and disintegration over a gigantic collision energy range is evident. Perhaps the most relevant discovery pertinent to this CPOD2016 conference was made by Marek in his detailed experimental study of the K$^+/\pi^+$ ratio~\cite{Gazdzicki:2010iv}. This ratio shows the so called \lq\lq Marek-Horn\rq\rq. At an early time in our discussions of the Horn, I told Marek that within our chemical non-equilibrium SHM approach I could certainly explain his data. I remember how he looked with both hope and skepticism at me saying \lq\lq please publish.\rq\rq\ The reason I readily made the claim is that the non-equilibrium SHM was created in order to describe QGP explosive disintegration and what else could Marek be observing?

This conversation took place mid-February 2003 as we drove together between a Winter school in Karpacz near Wroclaw and Frankfurt. We stopped for lunch and to explore Prague. When we returned after lunch to the car parked legally in front of the Charles Bridge police station we discovered to our amazement that someone managed to take out of the locked trunk a bag that contained an essential gift, a Polish bottle of \lq Wyborowa\rq. We soon replaced the bottle with \lq Beherowka\rq, and continued our discussion for a few enjoyable hours followed by even more wonderful hospitality offered by Magda at her home, before my flight the following morning back to the US. In retrospect it is clear that the good time I always had visiting with Marek prevented us from writing up the physics we discussed. Such is the cost of friendship.
 
Marek\rq s interest in seeing a theoretical and independent interpretation of his Horn-result was motivated by several loud and critical voices suggesting that more complete experimental results were needed before one could take the Horn seriously. I recall that all those saying this were also not in a position to come close to agreeing with \lq Marek\rq s-Horn\rq\ data. However, Jean Letessier and I demonstrated by April 2005~\cite{Letessier:2005qe} that these experimental data were within just about one s.\,d. consistent with the SHM chemical non-equilibrium model hadronization of QGP. 

The question that preoccupies is what if at all changes at the tip of the Horn estimated  to be near but below $\sqrt{s_\mathrm{NN}}=7$\,GeV. We saw that there is a change in the speed of increase of strangeness $s$ production yield as compared to the entropy $S$ production yield. Above the Horn-tip the ratio $s/S$ rises less rapidly compared to the pre-Horn energy behavior, see Fig.~29 in Ref\,.\cite{Rafelski:2015cxa}. In order to understand the situation, in Ref.\,\cite{Letessier:2005qe} we also studied the energy cost per strange quark pair. This value was obtained by evaluating the total strangeness yield of all produced hadrons along with the total final state thermal energy of the fireball as carried by the emitted hadronic particles. 

We found that as a function of collision energy above the Horn the cost in thermal energy per $s\bar s$-pair was much reduced as compared to the below-Horn evaluation. The summary of our results in SPS-RHIC domain is presented on the right in \rf{Kpi2008Fig}. While the transition was smooth, a distinct angle appeared, suggesting a transition between two different mechanisms of strangeness production in thermal hadron matter occurring near the Horn tip. This must be understood in the two-step mechanism context of strange hadron production: first strangeness is produced in microscopic reactions inside the hot fireball and second it is retained in hadrons produced during the hadronization process. 
 
\begin{figure}[bht]
\centerline{%
\includegraphics[width=4.8cm]{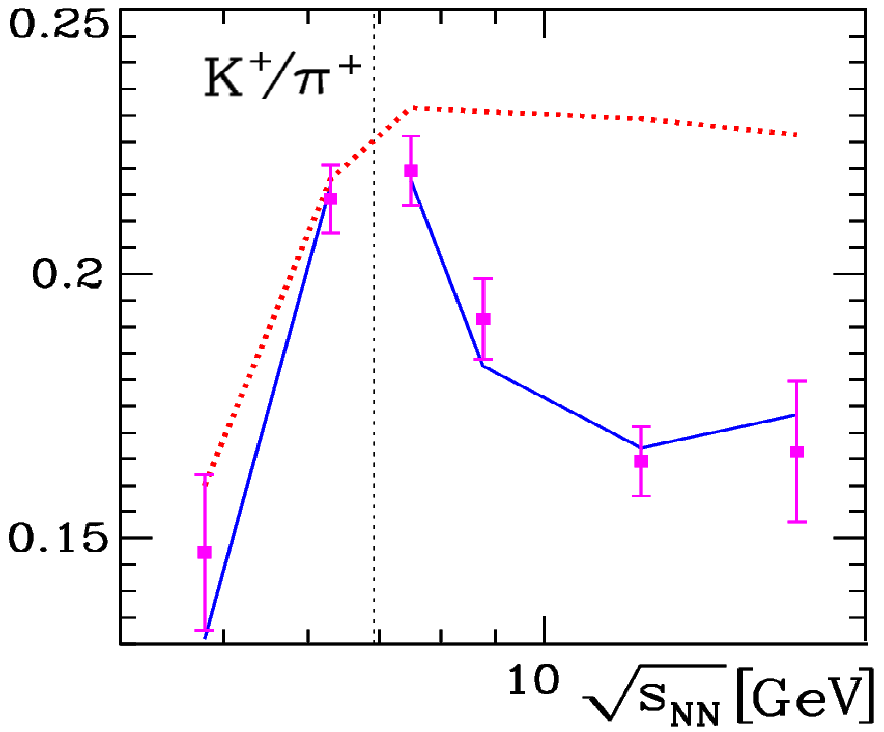} 
\includegraphics[width=7.0cm]{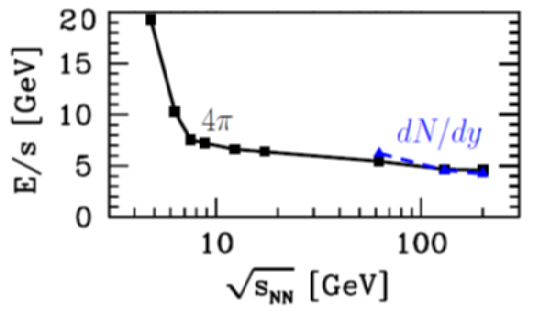}}
\caption{\underline{Left:} K$^+/\pi^+$-ratio experiment and theoretical fit adapted from 2009 data analysis, Ref.\,\cite{Rafelski:2009gu}; AGS (lowest $\sqrt{s_\mathrm{NN}}$) and NA49-SPS energy range. \underline{Right:} Cost in fireball thermal energy of a strangeness pair, $E/s$ as a function of CM collision energy $\sqrt{s_\mathrm{NN}}$. $4\pi$ results (black) are estimates for RHIC, line guides the eye; RHIC domain (blue) shows $(dE/dy)/(ds/dy)$. Update of result of Ref.\,\cite{Letessier:2005qe}.}
\label{Kpi2008Fig}
\end{figure}

Let me also clarify why considered as a function of collision energy, in the experimental data we see a dip in K$^+/\pi^+$ ratio above the Horn, even if it is \lq cheap\rq\ in terms of thermal energy, to make $s\bar s$-pairs. The changing chemical composition of the fireball disintegration products is responsible for this effect, with additional pions being produced by hadron resonances arising at lower chemical potential and higher temperature. It is notable  that at the upper end of RHIC energies and 20 times higher LHC energies, a similar K$^+/\pi^+$ ratio arises in agreement with all the interpretational remarks made here.

Our April 2005 analysis of the Marek-Horn~\cite{Letessier:2005qe} was developed after our SHM program suite called SHAREv1~\cite{Torrieri:2004zz} ({\bf S}tatistical {\bf H}adronization with {\bf RE}sonances)  was complete. SHAREv1 was created by two groups, Krak\'ow and Tucson, and it was properly debugged. Debugged means here that more than one group of researchers devote time and effort to query each other, trying a program in different contexts and insisting that the common product agrees to the last significant digit with the results obtained with prior SHM programs both groups had available, and that there is stability of the result for example when some data points are removed. 
 
In the Horn context it is interesting to study the RHI collision energy dependence of the hadronization condition  shown in the hadronization temperature -- chemical potential plane scatter plot in \rf{TEcol}, which is an update of Fig.~9 in Ref.\,\cite{Rafelski:2015cxa} (see there for all pertinent references to data and lattice QCD). All model values that we see in \rf{TEcol} well above the lattice value $T_c$ on the left margin in \rf{TEcol} are obtained either assuming full chemical equilibrium, or using ROOT where the hadronization contents comes from the  THERMUS~\cite{Wheaton:2011rw}, or in most if not all cases, both. We see in \rf{TEcol} that the full chemical non-equilibrium results obtained using SHARE are convincingly in temperature below the phase transformation boundary between QGP and hadron phase obtained in lattice-QCD. 
\begin{figure}[bht]
\centerline{%
\includegraphics[width=9.6cm]{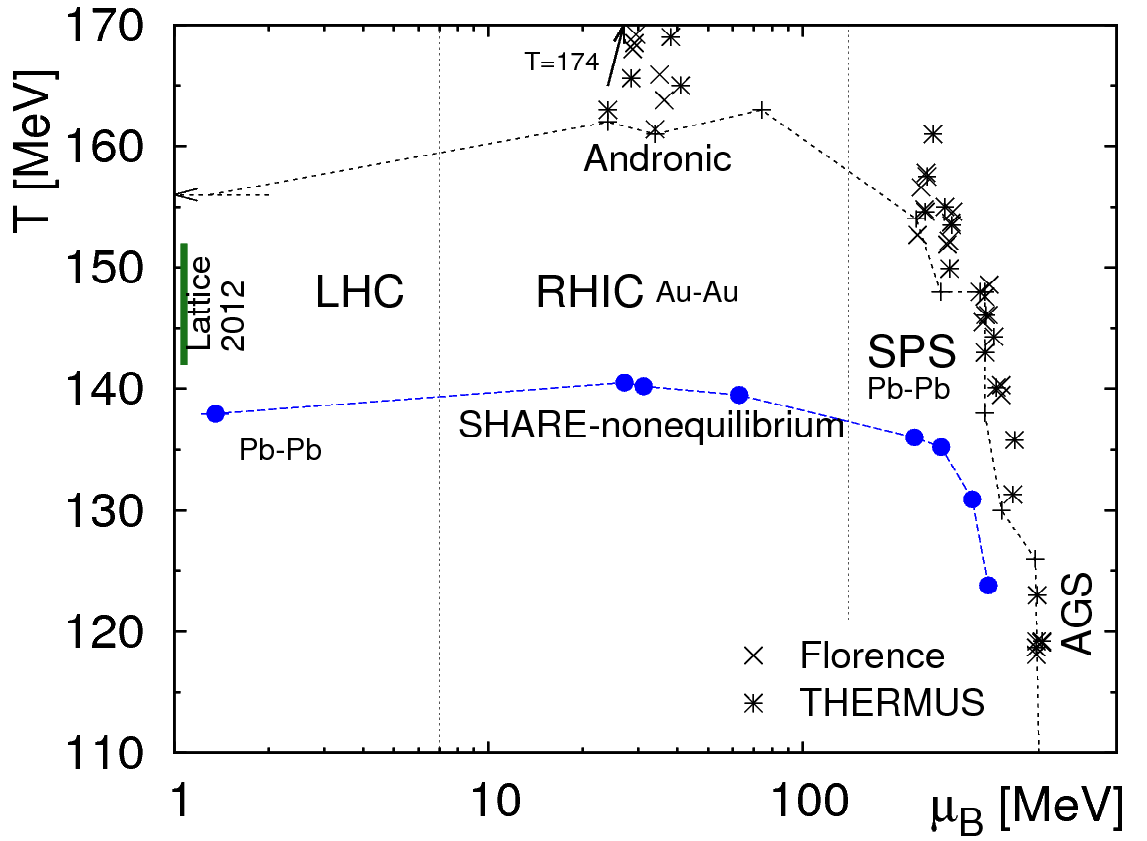}}
\caption{Update~\cite{Petran:2016} of results Fig.~9 in Ref.\,\cite{Rafelski:2015cxa} showing the $T,\mu_\mathrm{B}$ scatter diagram showing current lattice value of critical temperature $T_c$ (bar on left), as compared to SHM results of different groups for analysis performed for different collision energies as indicated. Our SHARE results are seen as full blue circles, dashed blue line guides the eye; the GSI-Andronic chemical semi-equilibrium results are crosses with dashed black line guiding the eye. Other results are also shown, including those obtained using ROOT platform, see text.}
\label{TEcol}
\end{figure}

The incoherent   hadronization condition results seen e.g. in \rf{TEcol} at the top edge  of the diagram are believed to be due to a ROOT analysis platform problem:  once one reads out the yield of any decaying hadron, say K$^*$, these particles are removed from the decay chain and do not contribute, in this example, through K$^*\to$ K+$\pi$, to the final kaon and pion yields. The impact on Kaon yields in this example case could be at 30\% and more. Hyperon yields can be also significantly impacted. Sometime in Fall 2011 -- I recall the year based on the soon after presented work Ref.\,\cite{Petran:2011aa}, together with Ph.D. student Michal Petran -- I had a meeting with a RHIC-STAR experimental group. They had analyzed their $\sqrt{s_{NN}}=62.4$ GeV  with a different physics outcome compared to us. 

As we were sipping coffee and looking at the differences in STAR and our analysis in detail I made a bid to convince the STAR group that there must be a programming error in their work. I was told that it is impossible that the ROOT platform used in their work is wrong, \lq everybody is using this\rq. As the popular sentiment against SHARE grew, I made my point about SHARE being right and ROOT being wrong in an open discussion at the SQM2013 Birmingham meeting, pointing out that the young students were not in a position to recognize a programming error. I believe that today we all know I was right. However, this matter deserves a further in-depth study which will establish once and for all a reliable  SHM analysis tool.

\section{Statistical Hadronization}\label{sec:SHM}
\subsection{Sudden hadronization}\label{ssec:fire}
In the absence of a QCD phase transition in a fully equilibrated system with physical flavor masses~\cite{Borsanyi:2016ksw}, the lattice QCD (L-QCD) method presents us the properties of a mixed phase containing presumably at the same time thermally equilibrated and changing in abundance as temperature decreases, quarks, gluons, and hadrons: we believe this since for high $T$ we find good agreement of L-QCD with perturbative results involving only quarks and gluons, and for small $T$ we find that L-QCD agrees with hadron gas limit, see Sect.\;4 in Ref.\,\cite{Ding:2015ona}. Thus as quark and gluon content is turned off, the hadron content is turned on. This is only a best guess; at this time we do not have access to chemical composition properties of fully equilibrated QGP, nor have we obtained from L-QCD information about QGP properties when some of its components are out of chemical abundance equilibrium.

However, the QGP fireball formed in laboratory is governed by a rapid dynamical evolution. We anticipate the fireball evolution time scale which is determined by the geometric size of the colliding nuclei, thus at the level of $\tau_f\simeq 10\;\mathrm{fm}/c$. In consideration of this time constant none of the stages of fireball evolution can possibly be well equilibrated. Upon first heavy-ion impact and ensuing collisions a dense partonic matter fireball is created which lacks both thermal and chemical equilibrium. The fireball internal energy content feeds the fireball expansion. Thermal equilibrium develops and one can now say that internal nearly thermal pressure of partons accelerates the fireball expansion. 

When the fireball temperature decreases below $T_\mathrm{q}\simeq 250$\,MeV we expect in the absence of a phase transition the formation of hadrons and creation of hadron-quark-gluon mixed phase. However, I do not believe that given the time constant of $\tau_\mathrm{qh}\simeq 3$\,fm/c that this phase lasts till the system reaches the pure hadron phase temperature $T_\mathrm{h}\lesssim 150$\,MeV; a mixed phase can develop. In other words, the greatest difference of laboratory dense matter experiments to L-QCD computations is the inability for mixed hadron with quark-gluon phase. 

Thus I believe that the quark-gluon composition of the QGP does not evolve smoothly into hadrons with the possible exception of heavy flavor $c,b$ seeded states~\cite{Thews:2000rj}. In a fireball that retains its quark-gluon composition down to temperatures that are near and even below $T_\mathrm{h}$, the chemical QGP composition is strongly out of chemical equilibrium. The proper technical statement is that the expanding laboratory fireball is \lq supercooling\rq. In such a situation we can encounter an effectively discontinuous \lq sudden\rq\ transformation akin to a phase transition. In this case the process of conversion, or more accurately put, fireball disintegration into individual hadrons, will be characterized by a universal transformation condition that reminds of effective 1st order transition. 

Should the temperature $T_\mathrm{h}$ be low enough, these hadrons are created free-streaming into vacuum. This is~\cite{Rafelski:2000by} \lq sudden hadronization\rq, used previously in data analysis~~\cite{Rafelski:1991rh}. The free-streaming hadrons in this case carry information allowing the study of the features of laboratory created QGP near to the breakup point, and through specific observables, also the integrated fireball history. At the time of writing no well-defined procedure to relate the outcome of this fireball hadronization analysis to L-QCD results is known. 

Developing models of heavy-ion formed QGP and the fireball breakup we leave the domain of exact computational L-QCD where well-defined parameters characterize numerical precision of results obtained. We enter the domain of models and physical hypothesis where agreement with experimental reality governs the understanding of the processes observed. Should L-QCD acquire the capability to study chemical nonequilibrium conditions, a closer link can be forged between laboratory experiments and L-QCD.

However, there are some constraints governing hadron production. In any hadronization process and thus also in context of the sudden hadronization the baryon number $B$ and the electric charge $Q=Ze$ content of QGP is transferred into the produced hadrons without any change. Entropy $S$ content can increase but since the multi-quark hadron states as compared to quarks carry less entropy, it is in general believed that hadronization struggles to transfer the entropy from quarks to hadrons and thus little if any additional entropy production in hadronization occurs. Similarly it is expected that no significant production or destruction of strangeness occurs during hadronization, this expectation is based on kinetic theory modeling of reaction rates. Because of the nature of the QGP breakup all hadron states can be produced subject to these constraints with equal probability in the sense that the phase space weight is all that controls the relative abundance. This is the principle governing the statistical hadronization model, SHM.

\subsection{Analysis of Experimental Data}\label{ssec:analysis}
The statistical hadronization model (SHM)  was born in 1990/91 when the first experimental results about the production of strange antibaryons became available and an analysis procedure was proposed~\cite{Rafelski:1991rh} to determine the magnitude of chemical potentials based on relative particle abundances. Since strangeness was known to equilibrate slower in the absence of gluons a chemical non-equilibrium parameter $\gamma_s$ describing the phase space occupancy of strangeness amongst hadrons was introduced. In following years the entropy conservation in hadronization process gained in importance, and in order to be able to achieve this the phase space occupancy of emerging light quarks $\gamma_q$ was introduced 1997/8~\cite{Letessier:1998sz}. 

The development of SHM for particle yield characterization was preceded by efforts to account more comprehensively also for the shape of particle spectra. In pursuit of this effort we recognized the model dependence that is introduced to model the explosive disintegration of the QGP fireball. This dynamical process influences the momentum spectra and is much more difficult to understand compared to the momentum integrated particle yields which are therefore our target of interest, being independent of how fireball matter expands. 

An analysis of experimental hadron yield results requires a significant bookkeeping effort, in order to allow for resonances, particle widths, full decay trees and isospin multiplet sub-states. We therefore have developed a special program package, SHARE ({\bf S}tatistical {\bf H}adronization with {\bf RE}sonances) which has seen three evolution stages and is available for public use~\cite{Torrieri:2004zz,Torrieri:2006xi,Petran:2013dva}. SHAREv3 (SHARE with CHARM) incorporates  more than 500 hadrons, updated last time according to the 2012 particle data group. The quark flavor chemistry is addressed in full as is necessary if one wishes to describe the totality of produced hadrons in a wide range of collision energy and centrality. Therefore we have available an extended set of chemical parameters motivated by the microscopic model of QGP hadronization including  $\gamma_i$, the phase space occupancy, allowing to fix the  the number of quark-antiquark pairs and thus the \lq absolute\rq\ chemical equilibrium.

Bulk matter constraints such as the vanishing of net strangeness $\langle s - \bar{s}\rangle=0$ are defaults in SHARE that can be relaxed as the user explores the parameter domain. SHARE allows to prescribe a mean charge per baryon (\ie\ for heaviest ions $\langle Z\rangle/\langle B\rangle\sim 0.39$) as SHARE has the SHM parameters allowing the count of charge and baryon number separately, a feature which is essential to address neutron-proton asymmetry.

The finite volume of hadrons plays an important role in the phenomenological study of properties and transformation of different phases of hadronic matter; a modern perspective on this topic is offered in Ref.\,\cite{Rafelski:2015cxa}. However, in the study of hadronization we can treat hadrons as point particles. Hadron proper volume plays an important role should hadrons decouple in conditions that are ultra-dense. We assumed that such a situation is physically unlikely since overlapping hadron volumes imply continued hadron scattering, whence we conclude that the idea to seek condition of free-streaming hadrons is incompatible with the idea that the finite size of these hadrons is of relevance. 

Considering that in our best fit the chemical freeze-out is occurring at relatively low particle and energy density we found  no need for finite hadron size. While we noted that for some data sets one can also find another ultra-high density fit, this solution usually disappears in a more detailed consideration where we require continuity of the results considered as a function of collision energy and geometric collision centrality. As discussed in Ref.\,\cite{Rafelski:2015cxa}, the effect of finite hadron volume and the small estimated effect of not-yet discovered and thus not-yet SHARE incorporated hadron resonances cancel each other.

Counting parameters and constraints we see that for NA61/SHINE one needs 7  parameters with two constraints  in chemical non-equilibrium SHM approach. Thus we need 5+ particle abundance yields to proceed. The number of degrees of freedom (dof=\#data +\#constraints -\#parameters) is in general not large -- however,  when 10 or more particle abundances are measured, we have considerable consistency verification and expect a fit that should have $\chi^2/\mathrm{dof}\simeq 0.6$. This value is well below unity since only in the limit of dof$\to\infty$ and statistical (Gaussian) error we expect $\chi^2/\mathrm{dof}\to 1$. 

Finding a best fit for NA61/SHINE may be to some a trying effort. This is so since we are searching for a sharp minimum of $\chi^2/\mathrm{dof}$ in a 7-dimensional space to the experimental hadron yield data and two constraints. The constraints are effectively additional data points in the fit. One usually runs a random search many times with many different initial departure points using different search algorithms. This procedure can be speeded up for QGP energy domain as we already have found  convergence to a physical fireball with universal \eg\ energy density, which is computed using all produced hadrons. Thus by introducing a loose constraint that the energy density is around $0.45\pm0.1\;\mathrm{GeV/fm}^3$ (or equivalent constraint for pressure $P=80\pm 5\;\mathrm{MeV/fm}^3$ more about this follows) we find a best fit more rapidly. 


\subsubsection{Results: Strangeness and Entropy in QGP}

All quark flavors can be produced in initial parton collisions. Strangeness differs from the heavier quarks by the relatively low mass production threshold. This means that it continues to be produced in ensuing in medium parton processes dominated by gluon fusion~\cite{Rafelski:1982pu}. This coupling to the gluon degree of freedom implies that the QGP abundance rises rapidly at first, but it can also fall, tracking the cooling of the gluon degrees of freedom. Ultimately, when parton temperature and density is sufficiently low, strangeness undergoes the chemical freeze-out process. 

In order to characterize the source of strange particles our target variable is the specific per entropy strangeness-flavor content $s/S$ which we want to track as a function of collision energy and centrality. Interpretation of the relation between strange antibaryon production and $s/S$ helps to understand the onset of deconfinement and the appearance of critical point.
Relative $s/S$ yield measures the number of active degrees of freedom and the degree of relaxation when strangeness production freezes-out. Perturbative expression in chemical equilibrium reads 
\begin{equation}
{s \over S}=\frac{{g_s\over 2\pi^2} T^3 (m_{ s}/T)^2K_2(m_{ s}/T)}
 {(g2\pi^2/ 45) T^3 +(g_s n_{\rm f}/6)\mu_q^2T}\simeq \frac{1}{35}\simeq 0.0286\;.
\end{equation}
When looking closer at this ratio one sees that much of ${\cal O}(\alpha_s)$ QCD interaction effect cancels out. However, for completeness we note that one could argue that $s/S|_{{\cal O}(\alpha_s)}\to 1/31=0.0323$. A stronger effect can be expected if we have QGP nonequilibrium
\begin{equation}
{s \over S}= { {0.03 \gamma_s^{\rm QGP}} \over 
 {0.4 \gamma_{\rm G} + 
 0.1 \gamma_s^{\rm QGP}\!\!+
 0.5\gamma_q^{\rm QGP}\!\! + 
 0.05 \gamma_q^{\rm QGP} (\ln \lambda_q)^2}}\to 0.03\gamma_s^{\rm QGP}\;.
\end{equation}
Finally, introducing the quantum statistics and doing numerical evaluation produces for $m_s=90$\;MeV the result seen on left in \rf{sSFig} where we also for comparison show this ratio computed in hadron gas. We see that equilibrated QGP is 50\% above equilibrated hadron gas. Actual strangeness production enhancement is larger considering that hadron gas governed reactions are further away from chemical equilibrium. 

We show on the right in \rf{sSFig} the centrality dependence for the ALICE and STAR 62 GeV $s/S$ results~\cite{Petran:2013lja}. The ALICE results show a quick rise to saturation in $s/S$ near to the perturbative QGP value shown on the left in \rf{sSFig}. This can be understood as an evidence that at time of fireball hadronization we study a fireball in which quarks and gluons (but not hadrons) are chemically equilibrated.

\begin{figure}[!ht]
\centering
\includegraphics[width=4.7cm]{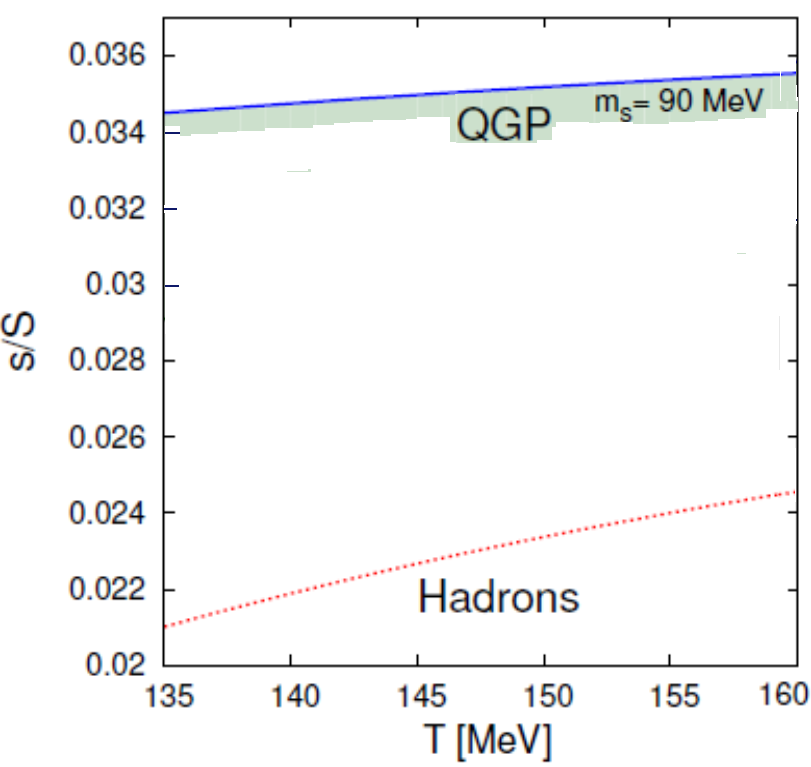}
\includegraphics[width=6.5cm]{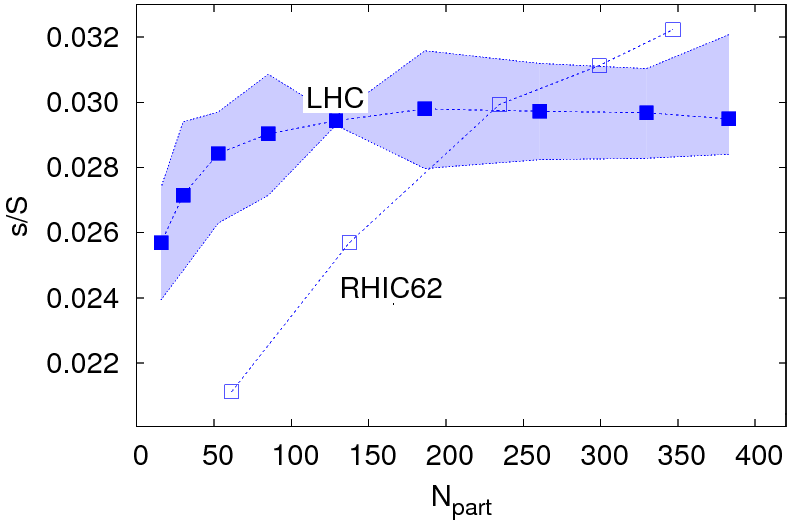}
\caption{Strangeness per entropy $s/S$: On left: as a function of temperature in QGP with $m_s=90$\;MeV and in the hadron resonance gas as defined by SHARE implemented mass spectrum. On right: Outcome of fit to ALICE $\sqrt{s_\mathrm{NN}}=2.76$\;TeV results as a function of centrality, expressed by the number of participants. Comparison with RHIC-62 GeV analysis (dotted line) based on STAR data which contains THERMUS-ROOT distortions.}
\label{sSFig}
\end{figure}

For the physical properties of the fireball at freeze-out we find the energy density $\varepsilon=0.45\pm 0.05$ GeV/fm$^3$, the pressure of $P=82\pm 2$ MeV/fm$^3$ and the entropy density of $\sigma=3.8\pm 0.3$ fm$^{-3}$ varying little as a function of reaction energy $\sqrt{s_\mathrm{NN}}$, collision centrality $N_\mathrm{part}$. This result was presented for SPS, RHIC and LHC energy domain in Refs.\,\cite{Rafelski:2009jr,Petran:2013qla,Rafelski:2014cqa}.

\subsubsection{Results: Hadronization Universality}

We make here the next step to understand hadronization universality~\cite{Petran:2016}. One can argue that the universality of freeze-out should be an invariant property not dependent on the observer frame of reference. Furthermore we are seeking a dimensionless quantity independent of scales. This leads us to explore as quantity characteristic for the universal hadronization the so called invariant measure which is obtained forming the trace of energy-momentum tensor normalized by $T^4$; that is 
\begin{equation}
I_\mathrm{m}=\displaystyle\frac{\varepsilon-3P}{T^4}\;.
\end{equation}
The result we find as a function of centrality based on analysis~\cite{Petran:2013lja,Petran:2013qla,Rafelski:2014cqa} is seen in \rf{TraceFig} on the left. What is striking is that a) The result is practically constant with value $I_\mathrm{m, L-QCD}\lesssim 5$ across all centrality, including the most peripheral collisions considered where strangeness abundance has not reached the full yield; and that the value $I_\mathrm{m, L-QCD}\lesssim 5$ is near to the peak value of $I_\mathrm{m, L-QCD}\lesssim 4.2$ obtained in L-QCD analysis~\cite{Ding:2015ona}. The  \lq experimental\rq\ value is, however, 20\% higher.  Note that we do not expect the fireball created in laboratory experiment to be in precise agreement with properties of a mixed hadron-quark phase that L-QCD represents.

\begin{figure}[!ht]
\centering
\includegraphics[width=12.cm]{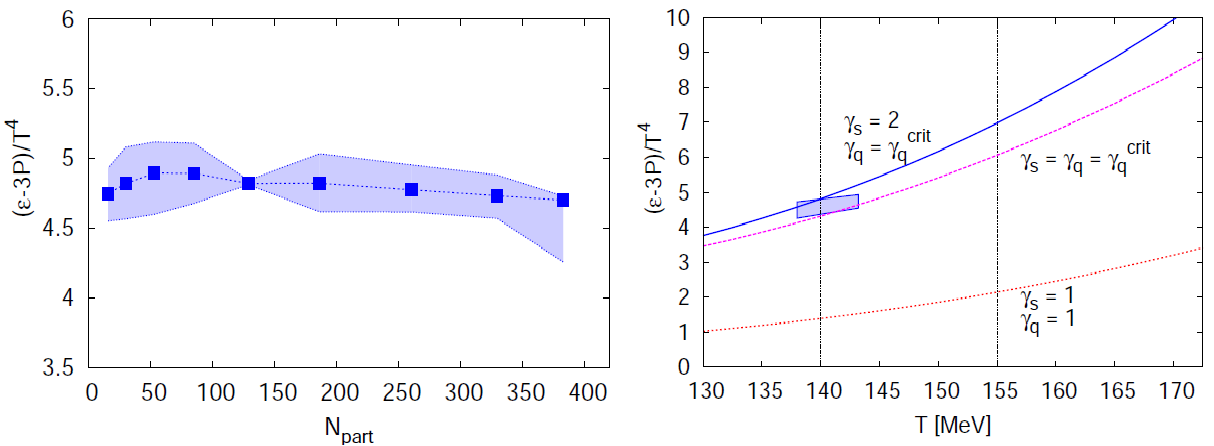}
\caption{Left: The trace anomaly measured in the study of the LHC QGP fireball, as a function of centrality (participant number); Right: trace anomaly in free-streaming hadrons with SHAREv3 mass spectrum; with domain of LHC analysis shown as blue area; the top solid line: a typical fit result; dashed violet line illustrates dependence on strangeness yield; bottom dotted line: equilibrium yield of hadrons~\cite{Petran:2016}.}
\label{TraceFig}
\end{figure}

In \rf{TraceFig} on the right we show how $I_\mathrm{m}$ depends on hadronization parameters, results are obtained using the SHAREv3 mass spectrum~\cite{Petran:2016}, while $I_\mathrm{m}$ increases with temperature, the high value of $I_\mathrm{m}$ we found is entirely driven by the value of the nonequilibrium parameter $\gamma_q$. The critical value $\gamma_q^\mathrm{crit}\sim 1.6$ used in \rf{TraceFig} on the right follows the value of temperature and is fixed so that the pion distribution is not condensing into a Bose condensate. We note small but relevant dependence on the value of $\gamma_s$.

Given the universal hadronization condition that we have obtained we believe that when the QGP hadronizes, it evaporates into free-streaming hadrons. There is no interlaced \lq phase\rq\ of hadrons, no afterburners are needed, except as we discussed to model light  nuclei abundances. 

\section{Conclusions}\label{sec:conc}
 
So here we are today looking back these many years. One can only ponder the many ups and downs! We learned a lot. We discovered a new phase of matter which \lq walks and quacks just like a QGP duck\rq. Comparing what we know today about the early Universe at high temperature with the physics context of 1980 one sees a complete change of paradigm. All this prompts the questions: How will the discovery chain continue? Which context touched by QGP research will evolve most in the next decades? Let me be speculatively optimistic and present an idea growing in my mind: by creating QGP we are learning how to convert (collision kinetic) energy into matter. I put forward that by the time Marek is 100 in 2056, we will have unraveled how to convert matter into energy.

By the year 2056 our present day students will have taken over our academic positions and be looking at their own retirements. That is a date beyond precise predictive power, my claim even if very speculative can stand. On the other hand perhaps we should look also at the near horizon. I hope that in these coming decades we will understand why QGP formation is so \lq easy\rq. We shoot two nuclei at each other and as it seems at energies we study in every collision we get QGP. That is a miracle. In early 1980\rq s I used to say -- please trigger on high multiplicity \ie\ high entropy production events and maybe you get enough counting rate to see QGP. Unraveling why we can unleash so easily the production of a high particle multiplicity/entropy is our priority today. I have worked on this in the past two years and have coined as explanation the concept of \lq acceleration frontier.\rq\ I believe that it is the large acceleration that prompts a collapse of quantum-vacuum structure and emergence of a high particle multiplicity and entropy. I will have more to say about this another time.

There is the related issue of the collision energy QGP formation threshold. On some days I believe that there is a threshold for QGP formation and Marek and NA61/SHINE will find it. But on other days I am not convinced that a true QGP production energy threshold exists. In this case the real question is how atomic nuclei avoid being in the quark matter form. Other people in past decades have also looked at this problem. Some concluded atomic nuclei are themselves in quark matter stage. There is a trail of work that sees nuclei as made of quarks and not nucleons, see for example Ref.\,\cite{Watson:1988up}. In such a case we are not liberating quarks in collisions but breaking the  QCD cluster bonds.  But that is again another story.

Marek and I have worked for 30 years to advance the study of strangeness production and QGP hadronization. We enjoyed finding  strangeness and antihyperons in large abundance.  We had a wonderful time discussing physics together. Because our interactions were enjoyable, we only wrote one paper together. But we impacted each other deeply with our ideas. This long-lasting dialog motivates me to laud Marek\rq s profound contribution to strangeness signature of QGP on the occasion of his 60th birthday!

\subsection*{Acknowledgments}
I acknowledge the contributions of Inga Kuznetsova, Jean Letessier, Michal Petran, and Giorgio Torrieri to the models and results presented here. I thank the organizers of CPOD for  very kind hospitality in Wroclaw.


\end{document}